\documentclass[11pt]{article}

\usepackage{geometry}

\usepackage{amsmath,amssymb,graphicx}
\usepackage{booktabs}
\usepackage[colorlinks=true,urlcolor=blue, linkcolor =blue,citecolor=blue]{hyperref}

\usepackage{algorithm,algorithmicx,algpseudocode}
\algblockdefx{MRepeat}{EndRepeat}{\textbf{Repeat}}{}
\algnotext{EndRepeat}

\allowdisplaybreaks

\title{\bf Inferring the heritability of bacterial traits in the era of machine learning}
\author{T. Tien Mai$^{(1),_{\footnote{Corresponding author. \texttt{the.t.mai@ntnu.no}}} } $ , 
	John A Lees$^{(5),(6)}$  , 
	Rebecca A Gladstone$^{(2)}$
	\\
	and Jukka Corander$^{(2),(3),(4)}$}

\date{
	{\footnotesize
		$^{(1)}$ Department of Mathematical Sciences,
		Norwegian University of Science and Technology, 
		Trondheim, Norway.
		\\
		$^{(2)}$ Department of Biostatistics, University of Oslo, 
		Oslo, Norway.
		\\
		$^{(3)}$ Department of Mathematics and Statistics, 
		University of Helsinki, Finland.
		\\
		$^{(4)}$ Pathogens and Microbes, 
		Wellcome Sanger Institute, Hinxton, UK.
		\\
		$^{(5)}$ European Molecular Biology Laboratory, 
		European Bioinformatics Institute EMBL-EBI, 
		Hinxton, UK.
		\\
		$^{(6)}$ MRC Centre for Global Infectious Disease Analysis, 
		Department of Infectious Disease Epidemiology, Imperial College, UK.
	}
}

\begin{document}
	\maketitle




\abstract{
Quantification of heritability is a fundamental desideratum in genetics, which allows an assessment of the contribution of additive genetic variation to the variability of a trait of interest. The traditional computational approaches for assessing the heritability of a trait have been developed in the field of quantitative genetics. However, the rise of modern population genomics with large sample sizes has led to the development of several new machine learning based approaches to inferring heritability. In this paper, we systematically summarize recent advances in machine learning which can be used to infer heritability. We focus on an application of these methods to bacterial genomes, where heritability plays a key role in understanding phenotypes such as antibiotic resistance and virulence, which are particularly important due to the rising frequency of antimicrobial resistance. By designing a heritability model incorporating realistic patterns of genome-wide linkage disequilibrium for a frequently recombining bacterial pathogen, we test the performance of a wide spectrum of different inference methods, including also GCTA. In addition to the synthetic data benchmark, we present a comparison of the methods for antibiotic resistance traits for multiple bacterial pathogens. Insights from the benchmarking and real data analyses indicate a highly variable performance of the different methods and suggest that heritability inference would likely benefit from tailoring of the methods to the specific genetic architecture of the target organism. 
}



\section{Introduction}
Heritability is a fundamental quantity in genetic applications \cite{falconer1960introduction,lynch1998genetics} which specifies the contribution of additive genetic factors to the variation of a phenotype. In the narrow-sense, heritability is defined as the proportion of the variance of a phenotype explained by the additive genetic factors. Heritability can be used to compare the relative importance between genes and environment to the variability of traits, within and across populations.  Together with GWAS (genome-wide association studies), the primary tool for discovering the genetic basis of a phenotype of interest,  heritability has been playing as a more and more critical role in exploring the genetic architecture of complex traits.

Current investigations of heritability in the quantitative genetics literature have focused on using the linear mixed-effect model framework \cite{speed2012improved,bulik2015ld,yang2010common,golan2014measuring,zhou2017unified,bonnet2016heritability,speed2017reevaluation}. In this framework, the effect sizes of genetic markers, usually SNPs (single nucleotide polymorphisms), are assumed to be independent and identically distributed random variables, and often the normal Gaussian distribution  is used for computational reasons. The genomic restricted maximum likelihood (GREML) and method of moments are the most widely used methods for heritability inference in this model, and some corresponding popular software are GCTA \cite{yang2011gcta}, LDSC \cite{bulik2015ld} and LDAK \cite{speed2019sumher,speed2012improved}. Although the linear mixed-effect model provides various ways to interpret correlations among covariates and traits, and is computationally tractable, it makes assumptions which do not necessarily accurately reflect the underlying genetics \cite{li2019reliable,lee2018accuracy,gorfine2017heritability,janson2017eigenprism,holmes2019summary,speed2017reevaluation,speed2019sumher}.  Some comparisons of different methods in this direction for estimating heritability have been recently conducted, for example, in \cite{zhou2017unified,evans2018comparison,weissbrod2018estimating,gorfine2017heritability,holmes2019summary}. 

The study of heritability estimation in the statistical machine learning community has been started relatively recently and it has not yet received wider attention. Current machine learning approaches often focus on the high-dimensional linear regression model where some sparsity regularizations could be used on the number of the covariates. This is a natural model for GWAS in modeling the whole-genome level contributions of genetic variation \cite{falconer1960introduction,lynch1998genetics}. The benefit of this model over the classical univariate approach in GWAS has been demonstrated for example in \cite{wu2009genome,brzyski2017controlling}. Several machine learning methods for making heritability inference have been studied:  a method of moments approach is proposed in \cite{dicker2014variance}; a convex optimization strategy is investigated in \cite{janson2017eigenprism} through a singular value decomposition;  maximum likelihood estimation is studied in \cite{dicker2016maximum};  some adaptive procedures have also been theoretically studied in \cite{verzelen2018adaptive}; two-step procedures based on high-dimensional regularized regression have been introduced in \cite{gorfine2017heritability,li2019reliable}; and,  a strategy for aggregating heritabilities through multiple sample splitting is introduced in \cite{mai2021boosting}. However,  up to our knowledge, a systematic numerical comparison of these different methods for estimating heritability has not yet been conducted. 

In this study, we provide a systematic summary of recent advances in machine learning methods for estimating heritability. More especially, we review and discuss the six above-mentioned different methods and compare them with GCTA method, a state-of-the-art method in quantitative genetics.  The application is carried in a bacterial GWAS context for estimating the heritability of antibiotic resistant phenotypes. While estimating heritability in human GWAS has been studied in numerous works, the topic has not yet been considered widely in bacteria, for the only prominent examples see \cite{lees2017genome,lees2020improved,mallawa2022heritability}. This is partly because bacterial GWAS poses unique challenges compared to studies with human or eukaryotic DNA in general, originating from highly structured populations and more limited recombination that produce in considerable linkage disequilibrium across whole chromosomes. 

Our article is structured as follows.  In Section \ref{sec_herit_method}, the problem of heritability estimation is introduced and a systematic review of machine learning methods for estimating heritability is given. In Section \ref{sc_data}, we briefly introduce the test datasets used in our evaluation.  Results and discussion of different methods on test datasets are presented in Section \ref{sc_result_discuss}.

\section{Heritability inference using machine learning methods}
\label{sec_herit_method}

The following notations are used in the work. The $\ell_{q , 0<q<+\infty}  $ norm of a vector $x \in \mathbb{R}^d$ is defined by $ \|x \|_q = (\sum_{i=1}^d |x_i|^q )^{1/q} $. For a matrix $A\in \mathbb{R}^{n\times m} $, $A_{i\cdot} $ denotes its $i-$th row and $A_{\cdot j} $ denotes its $j-$th column. For any index set $S \subseteq \{1,\ldots,d\} $, $x_S$ denotes the subvector of $x$ containing only the components indexed by $S$, and $A_S$ denotes the submatrix of $A$ forming by columns of $A$ indexed by $S$.

\subsection{The problem of heritability estimation}
Let $ y_i \in \mathbb{R} $ be the measured phenotype of subject $ i $ such that
\begin{align}
\label{linearmodel}
y_{i} = X_{i\cdot} \beta + \varepsilon_{i}, i = 1,\ldots,n
\end{align}
where  $ X_{i\cdot} = (X_{i1}, \ldots, X_{ip} ) \in \mathbb{R}^p $ is the vector of genotypes of subject $ i $ and $ p $ is the total number of variants; $ \varepsilon_1, \ldots, \varepsilon_n \in \mathbb{R} $ are unobserved independent and identically distributed (iid) errors with $ \mathbb{E} (\varepsilon_i) =0,  {\rm Var} (\varepsilon_i) = \sigma^2_\varepsilon >0  $; $ \beta = (\beta_1, \ldots, \beta_p )^\top \in \mathbb{R}^p $ is an unknown $p$-dimensional parameter.  We assume that $ X_{i\cdot}, i =1,\ldots,n  $  are iid random vectors and independent of $ \varepsilon_i $ with $ \mathbb{E} ( X_{i\cdot} ) =0$ and $ p\times p $ positive definite covariance matrix $ {\rm cov} ( X_{i\cdot} )= \Sigma  $.

Under the model \eqref{linearmodel},  for the $ i$-th observation it follows that
$
{\rm Var}(y_i) = {\rm Var}(X_{i\cdot}\beta ) + \sigma_{\varepsilon}^2 = \beta^\top \Sigma \beta + \sigma_{\varepsilon}^2 .
$
Our main focus is in estimating the narrow-sense heritability for the phenotype $y$ defined as
\begin{align}
\label{heritability:formula}
h^2 = \frac{ \beta^\top \Sigma \beta }{\beta^\top \Sigma \beta+ \sigma_{\varepsilon}^2 } .
\end{align}
In other words, this quantity computes the proportion of genetic differences present in the population variability of a trait.  It can benefit modeling the underlying genetic architecture of a trait, because a heritability close to zero means that environmental factors cause most of the variability of the trait, while a heritability close to 1 indicates that the variability of the trait is nearly solely caused by the differences in genetic factors.  

It is noted that as
$
\mathbb{E} [\| y\|_2^2/n ]   
= 
{\rm Var} (y),
$
one can use $\| y\|_2^2/n $ as an unbiased estimator for the denominator of the heritability.  Further,  \eqref{heritability:formula} can also be written as 
\begin{align}
\label{herit_thr_variance}
h^2 = 1 -  \frac{\sigma_{\varepsilon}^2 }{{\rm Var}(y)  } .
\end{align}
And thus,  an estimate of the noise-variance $ \hat{\sigma}_{\varepsilon}^2 $ (see e.g \cite{reid2016study}) can be used to estimate $h^2 $ rather than directly estimating the genetic variance $\beta^\top \Sigma \beta $.

Hereafter, we provide details for different methods for making heritability inference. We especially focus on methods that enable confidence intervals to be computed.
\subsection{Direct methods}

We first recall some methods that can directly estimate heritability from the data without identifying the genetic basis of the phenotype.

\subsubsection{Convex optimization approach}
Using a singular value decomposition (s.v.d.)  transformation, the work in \cite{janson2017eigenprism} proposes a method, called Eigenprism, to estimate the squared signal $ \beta^\top \Sigma \beta $ and thus heritability by solving a convex optimization problem. They also prove the asymptotic normality of their estimator.

More specifically, let $X = UDV^{\top} $ be a singular value decomposition (svd) and put $z = U^{\top}y $. Let $\lambda_{i, i = 1:n} $ denote the eigenvalues of $ XX^{\top}/p $. The authors of \cite{janson2017eigenprism} consider the following convex optimization problem, denoted by $ P_1 $,
\begin{align*}
\arg\min_{w \in \mathbb{R}^n} \max \left(\sum_{i=1}^n w_i^2 , \sum_{i=1}^n w_i^2 \lambda_i^2  \right) ,
\\
\text{ such that} \sum_{i=1}^n w_i = 0, \sum_{i=1}^n w_i \lambda_i =1.
\end{align*}
Let $ w^* $ be the solution to the problem $ P_1 $,  then the heritability estimator is given by
\begin{align*}
\hat{h}^2_{Eprism} = \frac{\sum_{i=1}^n w_i^* z_i^2 }{\| y\|_2^2/n   }    .
\end{align*}
With $ P^*_1 $ being the minimized objective function value, the $ (1-\alpha) $-confidence interval is given by
$
\left[  \hat{h}^2_{Eprism} \pm z_{1-\alpha/2} \sqrt{2 P^*_1} \right],
$
where $  z_{1-\alpha/2} $ is the $ (1-\alpha/2) $ quantile of the standard normal distribution.

The cost of this method stems mainly from the calculation of the s.v.d.  of $X$.  This will be expensive and slow for a genotype matrix with large dimensions. Moreover, we note that in practice the optimization in $ P_1 $ can fail sometimes and, in addition, this method only works with high-dimensional data where $p>n $.

\subsubsection{Maximum likelihood estimation} 
Another direct method for estimating heritability is based on using the maximum likelihood method.  In the paper \cite{dicker2016maximum}, the authors derive consistency and asymptotic normality of the maximum likelihood estimation (MLE) under additional Gaussian assumptions. More specifically, the maximum likelihood problem is defined as
\begin{align*}
(\hat{\eta}, \hat{\sigma}^2_{MLE}) 
= 
\arg\max_{\eta, \sigma^2}  
\left\lbrace
- \frac{\log (\sigma^2)}{2}  -  \frac{1}{2n} \log \det    \left( \frac{\eta}{p}XX^{\top} + \mathbf{I} \right) 
\right.   
\nonumber  
\\
\left.
-  \frac{1}{2\sigma^2  n}  y^{\top}\left( \frac{\eta}{p}XX^{\top} + \mathbf{I} \right)^{-1} y \right\rbrace
\end{align*}
and the heritability estimate is given by
\begin{align*}
\hat{h}^2_{MLE} = 1 - \frac{ \hat{\sigma}^2_{MLE} }{ {\rm Var} (y)}.
\end{align*}
The authors also studied the consistency and asymptotic normality of this MLE estimator.  As a result,  the $ (1-\alpha) $-confidence interval is given by
$
\left[  \hat{h}^2_{MLE} \pm z_{1-\alpha/2} /\sqrt{2n} \right],
$
where $  z_{1-\alpha/2} $ is the $ (1-\alpha/2) $ quantile of the standard normal distribution.

This method appears to be quite efficient computationally as it only requires handling a matrix inversion of dimension $ n\times n $.

\subsubsection{Moments method} 
Heritability estimation based on method-of-moment has been proposed and studied in \cite{dicker2014variance,verzelen2018adaptive}.
Several estimators have been proposed in these works. However, when the covariance matrix $ \Sigma $ is non-estimable or expensive to estimate (as often is the case in practice), the reference \cite{dicker2014variance} proposed an estimator as follows, with 
$
S = X^\top X /n  
,  
\hat{m}_1 =  {\rm trace} (S)/p
,
\hat{m}_2 = \frac{1}{p} {\rm trace} (S^2) - \frac{p}{n}\hat{m}_1^2,  
$
and put
\begin{align*}
& \tilde{\sigma}^2 
= 
\left(1 + \frac{p \hat{m}_1^2}{(n+1) \hat{m_2}}\right)\frac{ \| y\|^2}{n} 
-  \frac{\hat{m}_1 }{n(n+1)\hat{m}_2} \| X^\top y\|^2 ,
\\
& \tilde{\tau}^2 
=
- \frac{p \hat{m}_1^2}{(n+1) \hat{m}_2} \frac{ \| y\|^2}{n} 
+  \frac{\hat{m}_1 }{n(n+1)\hat{m}_2} \| X^\top y\|^2 ,
\end{align*}
and the heritability estimate is then
\begin{align*}
\hat{h}^2_{Moment}
=
\frac{\tilde{\tau}^2}{\tilde{\tau}^2 + \tilde{\sigma}^2} .
\end{align*}
The asymptotic properties  of the moment estimator were proven in \cite{dicker2014variance} and some non-asymptotic results were derived in \cite{verzelen2018adaptive}.  These results allow to obtain  the $ (1-\alpha) $-confidence interval approximately as
$
\left[  \hat{h}^2_{Moment} \pm \log(1/2) \sqrt{p}/n \right] .
$

It is noted that this method requires to compute a $ p\times p $ matrix $S$ which is very costly for data with large dimensions.

\subsection{Plug-in Lasso type approaches}

We now discuss some naive plug-in methods that are based on sparsity penalized regression methods. These methods typically assume that there is a small subset of biomarkers (in the genotype matrix) that will be important to the phenotype and influence its variability. 
\subsubsection{Scaled Lasso}
The paper \cite{verzelen2018adaptive} studied the problem of heritability estimation through using a variance estimation, see formula \eqref{herit_thr_variance}, in high-dimensional sparse regression from the scaled Lasso method \cite{sun2012scaled}. The scaled Lasso (also known as square-root Lasso) is defined as
\begin{align*}
(\hat{\beta}_{SL}, \hat{\sigma}_{SL} ) = \arg \min_{\beta, \sigma}
\frac{1}{2n\sigma} \| y - X\beta \|_2^2 + \frac{n\sigma}{2} + \lambda \| \beta \|_1,
\end{align*}
where  \(\lambda >0\) is the tuning parameter.

The heritability estimate is
\begin{align*}
\hat{h}^2_{Slasso} = 1 - \frac{ \hat{\sigma}^2_{SL} }{ {\rm Var} (y)},
\end{align*}
and its honest $ (1-\alpha) $-confidence interval, given in \cite{verzelen2018adaptive}, is given by
$
\left[  \hat{h}^2_{Slasso} \pm \log(1/2)( k\sqrt{p}/n + 1/\sqrt{n} ) \right] ,
$
where $ k := \| \hat{\beta}_{SL}\|_0 $ the number of non-zero components in $  \hat{\beta}_{SL}$. It is noted that this confidence interval is rather honest in the sense that its width tends to be quite large. A sharp confidence interval for this estimator has not yet been constructed.

We note that the scaled Lasso method shares the same spirit as Lasso which returns a very sparse model. Therefore, heritability estimation for a phenotype with a polygenic basis by this method tends to lead to underestimation.

\subsubsection{Elastic Net}
From \eqref{heritability:formula}, a direct heritability estimate can be obtained by using a Lasso type method.  More precisely, let $S = \left\lbrace j : \hat{\beta} \neq 0 \right\rbrace $ where $ \hat{\beta} $ is an estimate from a Lasso-type method, we can calculate the heritability as in equation \eqref{heritability:formula} with $ \hat{\Sigma}_S = X_S X_S^\top/(n-1) $,
\begin{align*}
\hat{h}^2 = \frac{  \hat{\beta}_S^\top \hat{\Sigma}_S  \hat{\beta}_S  }{ {\rm Var} (y) }.
\end{align*}
Here, we focus on using the Elastic Net estimate for the regression parameters: $ \hat{\beta} = \hat{\beta}_{Enet} $. The Elastic Net has been shown to be especially useful when the variables are dependent \cite{zou2005regularization} (LD structure), which is particularly relevant in bacterial genome data \cite{earle2016identifying}. The corresponding estimator is defined as
\begin{align*}
\hat{\beta}_{Enet} := \arg \min_{\beta} \frac{1}{n} \sum_{i=1}^{n}  \ell(y_i,\beta^T X_i) + \lambda\left[ \frac{(1-\alpha)}{2} ||\beta||_2^2 + \alpha ||\beta||_1\right],
\end{align*}
where  \(\ell( \cdot)\) is the negative log-likelihood for an observation.  Elastic Net is tuned by \(\alpha \in [0,1], \) that bridges the gap between Lasso (\(\alpha=1\)) and ridge regression (\(\alpha=0\)). The tuning parameter \(\lambda >0\) controls the overall strength of the penalty and can be chosen by using cross-validation. For example, 10-fold cross-validation is often used in practice. 

The paper \cite{lees2020improved} showed that Elastic Net  is  a promising method for bacterial GWAS data where the authors suggest using a small value of $ \alpha $, e.g $ \alpha = 0.01$.  However, we would like to note that the confidence interval for heritability estimated by this plug-in method is not available yet. It is worth noting that a GWAS analysis using high-dimensional sparse regression, such as the Elastic net discussed above, would already provide the estimated effect sizes corresponding to the selected covariates.  Therefore, estimating the heritability by using these effect sizes would bring insight on understanding both the genetic architecture as well as the genetic contribution to a trait.

\subsection{Boosting Heritability method}
\label{sec_boosther}

We now present some advanced approaches in machine learning for making inference about heritability.  The core idea of these types of methods is to combine a covariate selection step with an estimation step, known as the selective inference approach. The selection step aims at either reducing the dimension of the problem or removing irrelevant biomarkers, after which a heritability estimation step is applied.  These  approaches have been shown  not only to improve the computational aspects of the estimation procedure, but to also yield more accurate results.

In this approach, the original data $ (y,X) $ is randomly divided into two disjoint datasets $ (y^{(1)} , X^{(1)} ) $ and $ (y^{(2)} ,X^{(2)} ) $ with equal sample sizes.

The HERRA method introduced in \cite{gorfine2017heritability} is first based on a screening method (e.g. as in \cite{fan2008sure}) to reduce the number of covariates below the sample size.  With the remaining covariates, the data are randomly split  into two equally sized subsets.  Then, a Lasso-type estimator is employed on $ (y^{(1)} , X^{(1)} ) $ to select a small number of important variables. After that,  the authors use the least squares estimator on $ (y^{(2)} ,X^{(2)} ) $ with only the selected covariates (from the Lasso-type estimator)  to obtain an estimate of the noise-variance. The procedure is repeated where the role of $ (y^{(1)} , X^{(1)} ) $ and $ (y^{(2)} ,X^{(2)} ) $ is switched to obtain another estimate of the noise-variance. Finally, heritability is calculated as in the formula \eqref{herit_thr_variance} where the noise-variance is the mean of the two estimated noise-variances.

The work in \cite{li2019reliable} has also proposed a ``two-stage" approach with sample-splitting.  More particularly, the data is also randomly split into two disjoint datasets with equal sizes. On the first data set $ (y^{(1)} , X^{(1)} ) $, they use a sparse regularization method based on Elastic net to first reduce the model by selecting the relevant variables. Then, on the second data set $ (y^{(2)} ,X^{(2)} ) $,  only the selected variables are used to estimate the heritability through a method of moments approach \cite{dicker2014variance} or GCTA method.  

These post-selection approaches guarantee that sparse regularization and variance estimation are carried out on independent datasets and thus the heritability will not be overestimated. However, both methods in \cite{gorfine2017heritability,li2019reliable} heavily depend on the way data is split. One can avoid this dependence by performing the sample splitting and inference procedure many times (e.g. 100 times) and aggregating the corresponding results. This is to make sure that the different latent structures possibly residing in the sample are properly taken into account in both the selection and estimation steps.  This is the core idea of the ``Boosting heritability" strategy in \cite{mai2021boosting}.

\begin{algorithm}{}
	\caption{Boosting heritability \cite{mai2021boosting} }
	\begin{algorithmic}[0]
		\State \textbf {Step 0:}  Using a screening method, such as the sample correlation \cite{fan2008sure}, to remove 25\% least associated covariates. 
		\MRepeat $\, B$ times from step 1 to step 4,
		\State  \textbf{Step 1:} With the remaining covariates, divide the sample uniformly at random into two equal parts. 
		\State \textbf{Step 2:} On the first part of the data, use Elastic net to select the important covariates.
		\State \textbf{Step 3:} On the second subset with only selected covariates from Step 2, estimate the heritability by using \eqref{herit_thr_variance} where the noise-variance is estimated using the least square method.
		\State \textbf{Step 4:} Repeat Step 2 and Step 3 by changing the role of the first and second subset.
		\EndRepeat
		\State \textbf{Final} $\rightarrow $ The final heritability estimate is the mean of the estimated heritabilities at each repeat.          
	\end{algorithmic}
	\label{Boostingheritability}
\end{algorithm}

The work in \cite{mai2021boosting} recently introduces a generic strategy for heritability inference, termed as “Boosting Heritability”, which generalized the ideas from the post-selection approaches by \cite{gorfine2017heritability,li2019reliable}.  Boosting heritability, detailed in Algorithm \ref{Boostingheritability}, uses in particular a multiple sample splitting strategy which is shown to lead in general to a stable and more reliable heritability estimate.  

More importantly, this procedure also provides an informative interval of estimated heritabilities that shows the range of the target heritability would belong to. We call this interval a ``reliable" interval.

\begin{table}
	\small
	\caption{Summary of test datasets} 
	\begin{tabular}{ | p{0.8cm} | p{1.4cm} | p{1.8cm} | p{0.8cm} | p{1cm} | p{.8cm}  | }
		\hline 
		Dataset name	& bacteria & antibiotic resistant phenotype(s) to
		& No. of samples & No. of genetic features & Reference
		\\ \hline
		MA  & \textit{Streptococcus pneumoniae} & Penicillin & 603 & 89703 & \cite{croucher2015population}
		\\ \hline
		Maela & \textit{Streptococcus pneumoniae} & 
		Tetracycline,   Co-trimoxazole,  Penicillin
		&   3069 & 121014 & \cite{chewapreecha2014comprehensive,lees2016sequence}.
		\\ \hline
		E.coli  & \textit{Escherichia coli} & 
		Amoxicillin,	Cefotaxime,	 Ceftazidime,	Cefuroxime,	Ciprofloxacin,	 Gentamicin
		& 1509 & 121779 & \cite{kallonen2017systematic}
		\\ \hline
		NG  & \textit{Neisseria gonorrhoeae} & 
		Azithromycin,	Cefixime,	 Ciprofloxacin,	 Penicillin, Tetracycline
		& 1595 & 20486 &  \cite{schubert2019genome,unemo2016novel}
		\\  \hline
	\end{tabular}
	\label{tb_datasets}
\end{table}

\section{Test datasets}
\label{sc_data}

We investigate performance of the different methods using four public bacterial datasets suitable for GWAS and simulated phenotypes based on the genetic architecture present in the data. We focus on estimating heritability of the antibiotic resistance phenotypes. Table \ref{tb_datasets} summarizes our test datasets. The heritability of the antibiotic resistance phenotype is expected to be high, meaning that the variability stems primarily from the observed genetic differences among these bacteria. 

\subsection*{Streptococcus pneumoniae: MA data}

{\it Streptococcus pneumoniae} (the pneumococcus) is a common nasopharyngeal commensal that can cause invasive pneumococcal disease.  Here, we consider two datasets for this bacterium, abbreviated as the MA and Maela data.

The MA data set consists of 616 \textit{S. pneumoniae} genomes from isolates collected from healthy children in an asymptomatic nasopharyngeal colonisation survey in Massachusetts between 2001 and 2007. The genomic data and phenotypes are publicly available through the publication \cite{croucher2015population}. After initial data filtering using a minor allele frequency (5\%) and removing missing data greater than 10\%, we obtain a genotype matrix of 603 samples with 89703 SNPs \cite{chewapreecha2014comprehensive}. We consider resistance to penicillin antibiotics as the phenotype \cite{croucher2013population}. The genome-wide association studies of this phenotype were conducted in  \cite{chewapreecha2014comprehensive}. The results are given in Fig. \ref{fg_MAdata}.  It is noted that the main mechanism of resistance to penicillin can be explained by the causal SNPs in the penicillin binding proteins \textit{pbp2x}, \textit{pbp2b} and \textit{pbp1a}, see  \cite{chewapreecha2014comprehensive}.

\subsection*{Streptococcus pneumoniae: Maela data}
The Maela dataset is a large \textit{S. pneumoniae} dataset which consists of 3069 whole genomes produced from randomly selected isolates from a longitudinal nasopharyngeal colonization study of infants and a subset of their mothers, performed between 2007-2010 in a rural refugee camp on the Thailand-Myanmar border \cite{chewapreecha2014comprehensive,lees2016sequence}.  The genomic data and penicillin MICs are publicly available from \cite{chewapreecha2014comprehensive}. Using a minor allele frequency threshold (5\%) and removing missing data greater than 10\%, we obtain a genotype matrix with 121014 SNPs.

We use a continuous phenotype corresponding to the inhibition zone diameters measured in the lab. These inhibition zone diameters are in practice used to define whether a sample is 'Sensitive' or 'Resistant' to an antibiotic, for some antibiotics, an 'Intermediate' designation is also given which we treat as resistant. We consider resistances to three different antibiotics as the phenotypes: tetracycline, penicillin and co-trimoxazole. The results are given in Fig. \ref{fg_Maela}. The genetic loci associated with these antibiotic resistances have been examined in genome-wide association studies in \cite{lees2016sequence}. The tetracycline resistance is conferred by the \textit{tetM} gene and the co-trimoxazole resistance is conferred by the SNPs in the \textit{dyr} gene \cite{maskell2001multiple}.

\subsection*{Escherichia coli: E.coli data }
\textit{Escherichia coli} is a common coloniser of the human gut but is also a leading cause of blood stream infections, in which antibiotic resistance is increasing. The \textit{E.coli} data from \cite{kallonen2017systematic} consists of 1509 isolates from a systematic survey of blood stream infections conducted in England between 2001-2012 with an alignment of 121779 SNPs (after initial data filtering with a minor allele frequency threshold 5\% and removing missing data greater than 10\%).

We consider resistances to six different antibiotics as the phenotypes: amoxicillin,	cefotaxime,	 ceftazidime, cefuroxime, ciprofloxacin, gentamicin reported as categorical phenotypes 'resistant', 'intermediate', 'sensitive' as in  \cite{kallonen2017systematic}. The results are given in Table \ref{tb_Ecoli}.

\subsection*{Neisseria gonorrhoeae: NG data}
\textit{Neisseria gonorrhoeae} is a sexually transmitted pathogen in which antibiotic resistance is rapidly evolving, leading to multidrug resistance (MDR) and some extremely drug resistant (XDR) strains. The NG data has been analyzed in these studies  \cite{schubert2019genome,unemo2016novel,grad2016genomic}. These 1595 clinical samples were from surveillance in the USA (2000-2013), Canada (1989-2003, selected for decreased susceptibility to cephalosporin) and the UK (2004-2013). We obtain a genotype matrix with 20486 SNPs (after initial data filtering using a minor allele frequency threshold 5\% and removing missing data greater than 10\%). We consider resistances to five different antibiotics as the phenotypes: azithromycin, cefixime, ciprofloxacin, penicillin, tetracycline. The results are given in Table \ref{tb_NGdata}.

\section{Results and Discussion}
\label{sc_result_discuss}

\subsection{Simulations}
\subsubsection{Simulation settings}
\begin{figure}
	\centering
	\includegraphics[scale=.35]{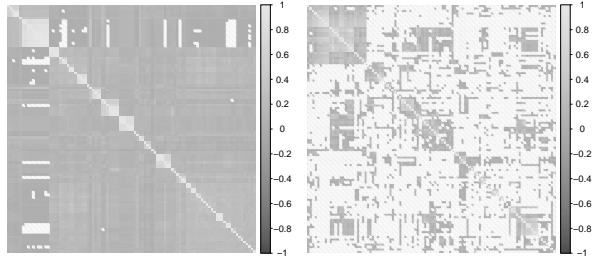}
	\caption{Sample covariance matrices of the 100 random SNPs (right) and 100 samples (left) in the genotype matrix shows the complex dependence structure present in the \textit{S. pneumoniae} Maela data.}
	\label{fg_covarmat}
\end{figure}

As the basis for systematically evaluating the performance of the different methods, we use a subset of the Maela data set (see Section \ref{sc_data})  to create a semi-synthetic dataset that incorporates levels of population structure and LD closely reflecting those present in natural populations (see Fig. \ref{fg_covarmat}).  This subset corresponds to a genotype matrix of 3051 samples and 5000 SNPs.  Using this real genotype matrix, we simulate the responses/phenotypes through the linear model defined in \eqref{linearmodel}.  For choosing the causal SNPs (non-zero effect sizes), we follow the penicillin resistance-like setting \cite{lees2016sequence,dewe2019genomic}: to select all SNPs from 3 genes (\textit{pbpX,pbp1A,penA}) as causal.

Given the chosen SNPs, regression coefficients $\beta^0$ are either drawn from the normal distribution $\mathcal{N}(0,1)$ or Student $ t_3 $ distribution. Because the true covariance of the genotype matrix is not given, we need to re-normalize the coefficient $\beta^0$ as $\beta = \beta^0  \sqrt{\sigma^2_\varepsilon  h^2/ (\beta^{0\top} \bar{\Sigma} \beta^0 ( 1-h^2 )) } $ to assure that the true corresponding heritability is approximating our target. Here $h^2 = 0.8 $ is our target heritability and $\bar{\Sigma} $ is the sample covariance matrix of the genotype matrix and the noise variance is fixed as $ \sigma_{\varepsilon}^2 = 1 $.

In simulations, the true covariance matrix of the genotype matrix is unknown, so phenotypes are approximated based on a given heritability using model \eqref{linearmodel}. To establish a benchmark for comparison, the noise variance is set to $ \sigma_{\varepsilon}^2 = 1 $, and an estimator is calculated using formula \eqref{herit_thr_variance}. This estimator, denoted by "oracle", is based on the true simulated values and cannot be used with real data. The methods are tested using the following settings: 'wholegenes' (whole genome analysis), 'subsample1500' (1500 randomly selected samples), 'subsample500' (500 randomly selected samples), 'causalgenes' (only true causal genes), and 't-effect' (effect sizes simulated from Student $ t_3 $ distribution). Additionally, the GCTA (mixed-effect) model is used to simulate phenotypes and is denoted as 'GCTA.model'. Results from 50 replications are shown in Fig. \ref{fg_simu}.

\subsubsection{Simulation results}

\begin{figure*}[!t]
	\centering
	\includegraphics[scale=.65]{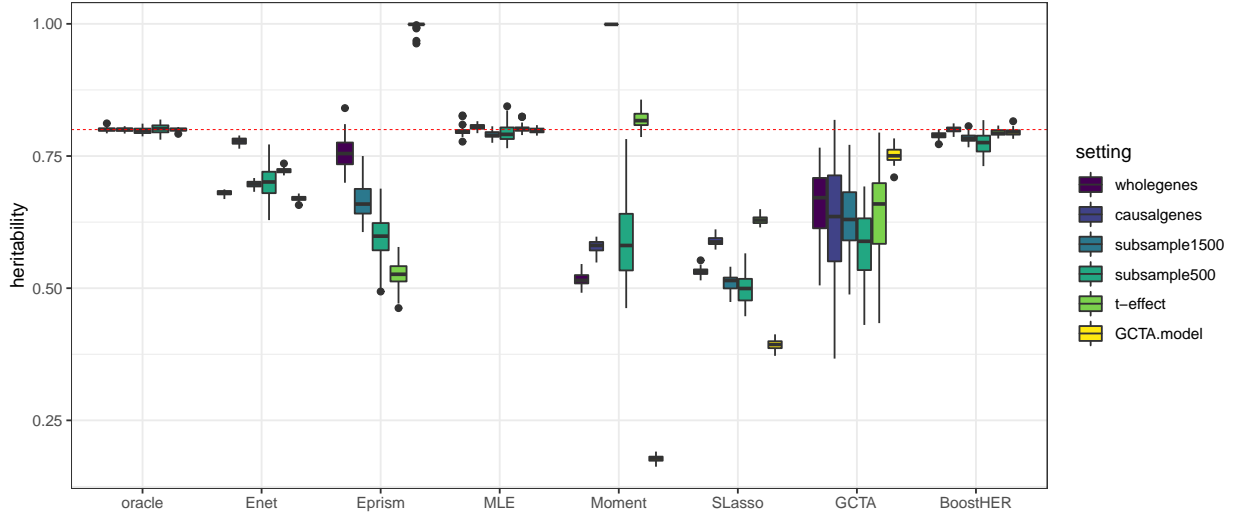}
	\caption{ Simulation results over 50 data replicates with Penicillin-like setting in Maela data, the true heritability is 0.8 (red-dashed line). Settings: 
		`wholegenes': run on whole genomes,
		`causalgenes': run only with true causal genes,
		`subsample1500': run with 1500 randomly selected samples, 
		`subsample500': run with 500 randomly selected samples, 		
		`t-effect': the effect sizes are simulated from Student $ t_3 $ distribution,
		`GCTA.model': the phenotypes are simulated from the GCTA model. }
	\label{fg_simu}
\end{figure*}

As seen in the simulation results presented in Fig.~\ref{fg_simu}, the 'oracle' estimator consistently demonstrates the highest level of accuracy and therefore serves as an effective benchmark for comparison. It is also evident that both the Elastic net and Lasso methods tend to underestimate the target heritability. However, the Elastic net method does provide a more reliable lower bound for the underlying heritability. The moment method, on the other hand, is found to be particularly unreliable in this context, failing to produce acceptable results across all settings. The Eprism approach, which is based on convex optimization, is also observed to be quite unstable in its performance.

On the other hand, both the Maximum Likelihood Estimation (MLE) and Boosting methods consistently provide accurate approximations of the target heritability for bacterial genome data across all the simulation settings that were considered. The GCTA method, however, is observed to consistently underestimate the target heritability, with a high degree of variability. This may be due to the fact that the GCTA method was specifically designed for mixed-effect models with different underlying assumptions about the effect sizes.

It has been observed that using the Elastic net (Enet) method to estimate heritability provides a reliable lower bound for the heritability and serves as a benchmark for comparison. However, studies such as \cite{qian2020fast,mai2021boosting} have shown that Enet tends to underestimate the true heritability due to a bias known to affect Lasso-type approaches. This is because coefficients associated with weak effects are shifted towards zero, even though these weak effects may still play a significant role in the overall genetic variation of a trait. On the other hand, the scaled Lasso (SLasso) approach often results in even lower heritability estimates than Enet. This is because it only selects a small number of covariates, resulting in a model that can only explain a limited amount of variation in the phenotype.

\subsection{Results with real data}

\begin{figure}
	\centering
	\includegraphics[scale=.36]{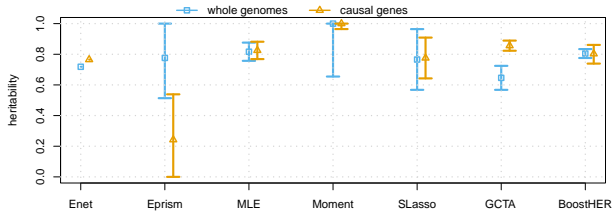}
	\caption{ Heritability estimation of antibiotic resistance in MA data.} 
	\label{fg_MAdata}
\end{figure}

\begin{figure}
	\centering
	\includegraphics[scale=.35]{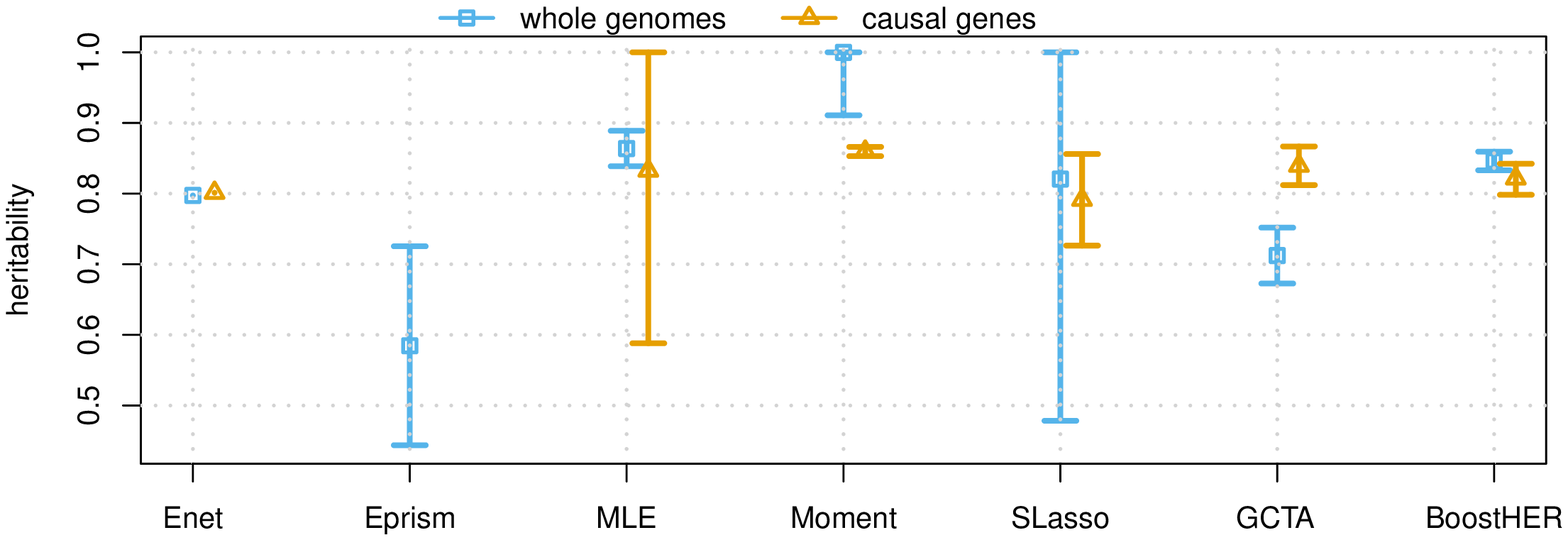}
	\includegraphics[scale=.35]{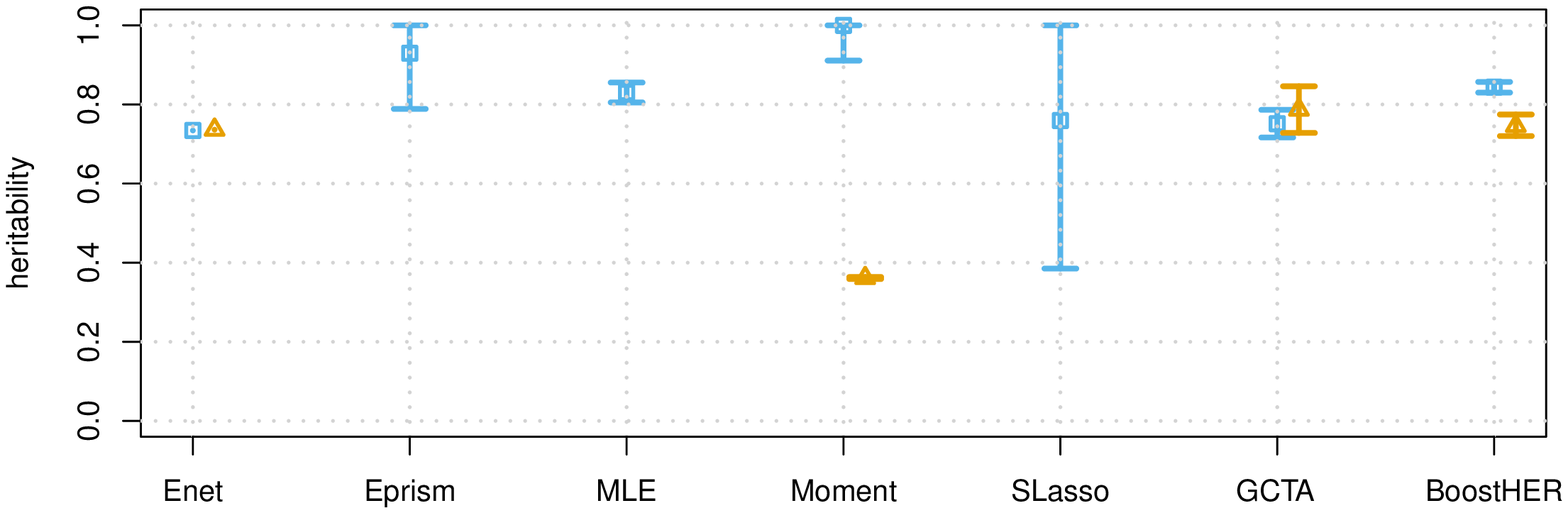}
	\includegraphics[scale=.35]{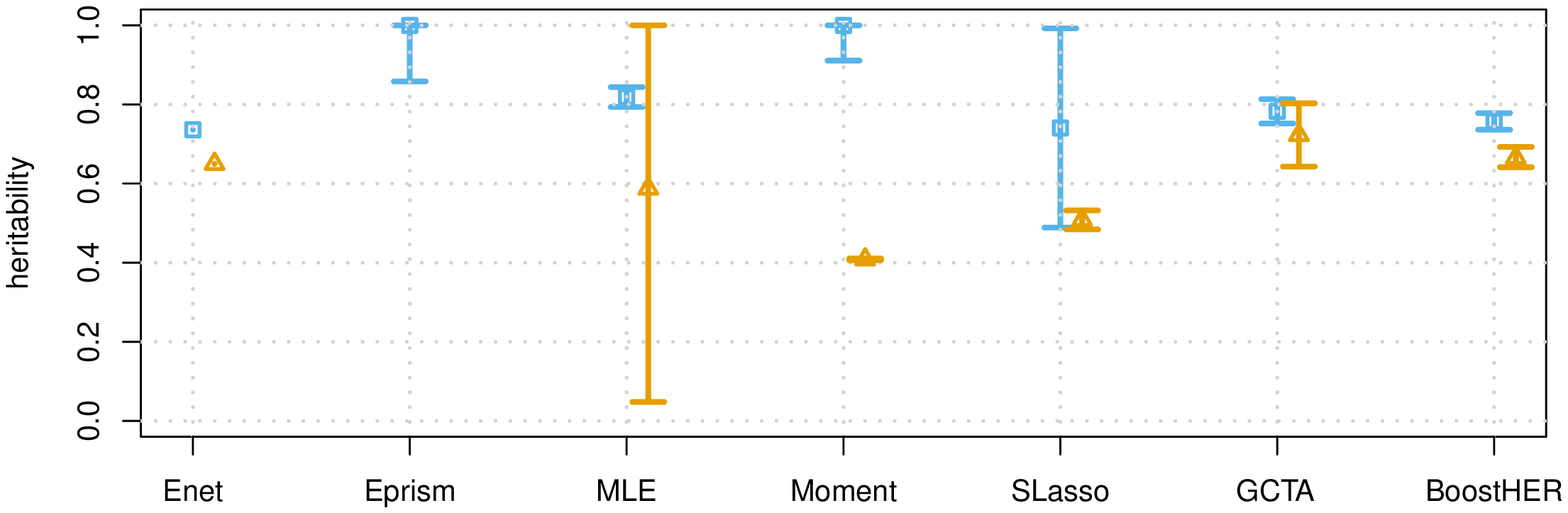}
	\caption{ Heritability estimations of antibiotic resistances in Maela data. The top plot is for Penicillin, the middle is for Tetracycline and the bottom plot is for Co-trimoxazole.} 
	\label{fg_Maela}
\end{figure}

The results on real datasets are given in Fig. \ref{fg_MAdata}, \ref{fg_Maela} and Table \ref{tb_NGdata}, \ref{tb_Ecoli}.

Overall, we see that direct approaches like convex optimization (Eprism) and moment method (Moment) are not able to deal with these data and return unstable results quite often.  On the other hand, the maximum likelihood method (MLE) yields consistent results in line with the outcome from Elastic Net (Enet) and GCTA methods.

The approach by combining selection and estimation steps together with multiple sample splitting as done in the Boosting heritability (BoostHer) method always returns reliable results. More specifically, its results are always at a higher value than the lower bounds given by the Enet method, and it can still work well when either MLE or GCTA method fails, as seen from Table \ref{tb_Ecoli}.

We also consider the estimation with respect to the causal genes for MA and Maela data. We observe that for MA data the estimation of heritability with only causal genes is slightly improved, see Fig. \ref{fg_MAdata}. While considering causal genes for Maela data does not gain an improvement, and in the case of Co-trimozaxole, it even lower downs the estimation, see Fig \ref{fg_Maela}. Note that this is desirable, since in practice we may not know all the causal genes, and thus we would like to obtain good results from using all predictors.

Regarding the uncertainty quantification, we can see that the confidence intervals (CIs) of MLE and the "reliable" intervals of BoostHer are stable.  More particularly, their widths are similar to those from the GCTA method.  In contrast,  other methods come with wider intervals and thus they can be harder to interpret. As an example, the confidence intervals for penicillin resistance heritability in Maela data (Fig. \ref{fg_Maela}) are as follows: the width for CI of GCTA is 7.91\%; of MLE is 5.02\%; and of BoostHer is 2.64\%; while the width of CI of Eprism is 28.16\% and of SLasso is 52.16\%. Since heritability is between 0 and 1, the latter two intervals are of limited value for interpretation.

\begin{table}[!ht]
	\scriptsize
	\centering
	\caption{Running times of different methods on 4 real datasets in seconds.} 
	\begin{tabular}{|  l | r r r r r r r| }
		\hline  
		data	& Enet	& Eprism & MLE	& Moment & SLasso	 & GCTA 		& BoostHER
		\\ \hline
		MA  & 26.1 &  38.2 & 0.8 & 162.7 & 10 & 0.3 & 82.0
		\\
		Maela  & 171.6 & 288.4 & 16.8 &  323.8 &  86.5 & 17.6 & 410.7
		\\
		E.coli  &  82.8 & 68.8 & 2.1 &  477.3 &  33.8 & 0.8 & 119.9
		\\
		NG  & 18.1 & 33.4 & 2.6 &  8.7 &  5.6 &  3.4 &  64.4
		\\ 
		\hline
	\end{tabular}
	\label{tb_runningtime}
\end{table}

\subsubsection*{Running time}
The indicative running times of all considered methods on four tested datasets are given in Table \ref{tb_runningtime}. The codes were executed on a Linux Redhat 64-bit operating system using a CPU with Intel-E7-4850v3 processor and 3TB of RAM, with the splitting step utilizing 10 CPU cores for parallelization. Overall, the maximum likelihood (MLE) method is the fastest method and also returns trustworthy results. The moment method seems to be computationally expensive, while its results are highly unstable and un-reliable.

\section{Conclusions}
In this study, we have conducted a thorough examination of multiple techniques for estimating heritability in bacteria, focusing on their precision and calibration of uncertainty. We have compared a diverse set of methods, including both traditional and newer techniques. Our findings revealed that, as anticipated, the maximum likelihood method is the fastest of all the methods we evaluated, while the method of moments is generally the slowest. However, it is important to note that none of the methods we tested had running times that would be considered prohibitive for practical use.

In our simulations, the maximum likelihood method demonstrated consistently strong performance. However, when applied to real data cases, its behavior was more mixed. This is likely caused by sensitivity to model assumptions, which tend to lead to a lack of robustness of MLE in general when the data deviate from these assumptions.

Our analysis also revealed that certain methods, such as Eprism, method of moments, and SLasso, consistently displayed poor performance and are not recommended for estimating heritability in bacteria. Conversely, the multiple sample splitting technique used in the BoostHer method emerged as the most reliable and accurate approach in our experiments. Overall, our findings provide useful insights for researchers and practitioners looking to determine heritability in bacteria.

It is important to note that our findings have implications for the estimation of heritability in other organisms as well. The patterns of linkage disequilibrium (LD), population structure, and existing studies of heritability are all quite different in humans and other organisms compared to bacteria. Therefore, it is crucial to consider the unique characteristics of the organism and the data when choosing a method for heritability inference. Furthermore, it is worth noting that the results of our computational experiments may not generalize to all possible scenarios and further research is needed to fully understand the applicability of these methods in different contexts.

Another investigation of heritability estimation for bacteria recently appeared, where linear mixed models were compared with the Elastic net and LD-score regression \cite{mallawa2022heritability}. The study found that linear mixed models showed poor correlation with the ground truth and typically overestimated heritability to a large degree, while Elastic net and LD-score regression methods were found to perform well. This observation is consistent with our findings, where we found that multiple sample splitting as in BoostHer appears overall as the most reliable and accurate approach in our experiments. It is worth noting that this recent study highlights the need for further research in the field of heritability estimation for bacteria, as it is clear that there is a lack of consensus on the best approach to use. The combined results of our study and the aforementioned study call for more research in the field of heritability estimation in bacteria to facilitate future studies of genetic architectures in bacteria.

\section*{Data Availability Statement}
The R codes and data used in the numerical experiments are available at:  
\url{https://github.com/tienmt/her_MLs} .

\begin{table*}
	\centering
	\scriptsize
	\caption{ Heritability estimation of antibiotic resistances in NG data (confidence intervals are in parentheses).} 
		\begin{tabular}{ | l | c c c c c |}
			\hline 
& Azithromycin	& Cefixime	& Ciprofloxacin	& Penicillin	& Tetracycline
			\\ \hline
			Enet & 0.69 & 0.78 & 0.91 & 0.68 & 0.73
			\\
			Eprism & 0.99 (0.82, 0.99) & 0.74 (0.56, 0.92) & 0.28 (0.10, 0.45) & 0.99 (0.82, 0.99) &  0.99 (0.82, 0.99) 
			\\
			MLE & 0.80 (0.76, 0.83) & 0.87 (0.83, 0.91) & 0.98 (0.94, 0.99) & 0.80 (0.76, 0.83) & 0.85 (0.81, 0.89)
			\\
			Moment & 0.08 (0.01, 0.14) & 0.99 (0.93, 0.99) & 0.19 (0.12, 0.26) & 0.06 (0.00, 0.12) & 0.58 (0.52, 0.65)
			\\
			SLasso & 0.33 (0.25, 0.40) & 0.69 (0.51, 0.86) & 0.95 (0.72, 0.99) & 0.44 (0.33, 0.55) & 0.37 (0.29, 0.45)
			\\
			GCTA & 0.81 (0.77, 0.85) & 0.77 (0.73, 0.82) & 0.85 (0.82, 0.88) & 0.73 (0.68, 0.78) & 0.81 (0.77, 0.84)
			\\
			BoostHER &  0.70 (0.66, 0.74) & 0.84 (0.81, 0.85) & 0.96 (0.95, 0.97) & 0.77 (0.74, 0.79) & 0.81 (0.79, 0.83)
			\\ \hline
		\end{tabular}
	\label{tb_NGdata}
\end{table*}

\begin{table*}
	\scriptsize
	\centering
	\caption{Heritability estimation of antibiotic resistances in E.coli data (confidence intervals are in parentheses).} 
		\begin{tabular}{ | l | c c c c c c |}
			  \hline 
			& Amoxicillin	& Cefotaxime	& Ceftazidime	& Cefuroxime	& Ciprofloxacin	& Gentamicin
			\\ \hline
			Enet &  0.56 &  0.44  & 0.26 
			&  0.23 & 0.75 &  0.31
			\\
			Eprism & 0.99 (0.76, 1.00) & failed & failed 
			& 0.55 (0.34,  0.77) & failed  & failed 
			\\
			MLE & 0.00 (0.00,  0.04) & 0.44 (0.40, 0.47) & 0.34 (0.30, 0.37) 
			& 0.32 (0.28,  0.35) & 0.90 (0.86,  0.94) &  0.34 (0.31, 0.38)
			\\
			Moment & 0.77 (0.55, 0.99) & 0.03 (0.00, 0.20) & 0.03 (0.00, 0.20) 
			& 0.12 (0.00,  0.28) & 0.10 (0.00, 0.26) & 0.05 (0.00, 0.21)
			\\
			SLasso & 0.42 (0.15, 0.70) & 0.17 (0.00, 0.36) & 0.14 (0.02,  0.26) 
			& 0.11 (0.03,  0.19) & 0.80 (0.49, 1.00) & 0.13 (0.03,  0.22)
			\\
			GCTA & 0.82 (0.77, 0.86) & 0.72 (0.65, 0.79) & 0.69 (0.62, 0.76) 
			& 0.31 (0.22,  0.41) & failed & 0.41 (0.31, 0.51)
			\\
			BoostHER & 0.67 (0.64, 0.70) & 0.53 (0.47,  0.58) & 0.42 (0.33, 0.51)
			& 0.40 (0.34, 0.45) & 0.89 (0.87,  0.92) & 0.45 (0.39, 0.50)
			\\  \hline 			
		\end{tabular}
	\label{tb_Ecoli}
\end{table*}

\section*{Competing interests}
No competing interest is declared.

\section*{Author contributions statement}
T.T.M.: conceptualization, formal analysis, write, review and edit the draft; J.A.L. and J.C.: write, review and edit the draft. R.A.G: data curation. All the authors read and approved the final manuscript.

\section*{Acknowledgments}
The authors would like to thank the editor and three anonymous referees who kindly reviewed the earlier version of this manuscript and provided valuable suggestions and enlightening comments.
This research was supported by the European Research Council (SCARABEE, no. 742158). TTM is supported by the Norwegian Research Council grant number 309960 through the Centre for Geophysical Forecasting at NTNU. JAL acknowledges funding from the MRC Centre for Global Infectious Disease Analysis (MR/R015600/1), jointly funded by the UK Medical Research Council (MRC) and the UK Foreign, Commonwealth \& Development Office (FCDO), under the MRC/FCDO Concordat agreement and is also part of the EDCTP2 programme supported by the European Union.

\bibliographystyle{apalike}

\end{document}